\documentclass[prl,aps,tightenlines]{revtex4}

\begin{document}

\title{History dependence of peak effect in CeRu$_2$ and V$_3$Si:
an analogy with the random field Ising systems}
\author{Sujeet Chaudhary,
A. K. Rajarajan,
Kanwal Jeet Singh,
S. B. Roy, and
P. Chaddah}
\address{Low Temperature Physics Laboratory,\\
Centre for Advanced Technology,\\ Indore 452013, India}

\begin{abstract}

\begin{center} \large\bf Abstract \end{center}

We present results of transport measurements showing distinct
path dependence of the electrical resistance in the
superconducting vortex state of single crystal samples of
CeRu$_2$ and V$_3$Si. Resistance measured in the vortex state of
both the systems prepared by field cooling (FC), indicates a
relatively higher degree of disorder than when it is prepared by
isothermal variation of field. Small oscillations of magnetic
field modify the resistance in the FC state, highlighting the
metastable nature of that state.  An analogy is drawn with the
FC state of the random-field Ising systems.
\end{abstract}                          
\maketitle

\section{Introduction}

Non-equilibrium properties associated with the peak-effect (PE)
in the C15-Laves phase superconductor CeRu$_2$ have drawn some
attention in recent years \cite{1,2,3}. It is observed that in
certain field (H) -- temperature (T) regime (which includes part
of the PE regime), the field-cooled (FC) vortex state in
CeRu$_2$ is more disordered than the vortex state obtained by
isothermal variation of H after zero field cooling (ZFC). This
FC vortex state in CeRu$_2$ is quite metastable in nature and
susceptible to external fluctuations \cite{2}.  The origin of PE
and the associated non-equilibrium properties in CeRu$_2$ remain
a matter of debate and various possibilities, starting from a
underlying phase transition\cite{1,4,5,6,7} to
dynamical crossover in flux-pinning properties \cite{8,9},
have been put forward.
 
In a recent communication \cite{3} we have drawn analogy between
the ZFC/FC response of the vortex state in and around the PE
regime of CeRu$_2$ and the ZFC/FC response of a random field
Ising systems (RFIM).  Through dc magnetization study in
polycrystalline samples of CeRu$_2$, we had shown that the
magnetization hysteresis in the FC state is distinctly more than
the ZFC state. This relatively large hysteresis in the FC state
suggested that the FC state was more disordered. It should be
recalled here that in the RFIM systems while there exists a long
range order in the ZFC state, the same is absent in the FC
state; instead the FC state is frozen into a disordered domain
structure, which is quite metastable in nature \cite{10,11}.
The analogous behaviour (if not the exact similarity) of these
two disparate class of systems has motivated us to investigate
more in this direction. In this communication we shall present
results of transport properties measured on good quality single
crystals of CeRu$_2$ and V$_3$Si. While the present results on
CeRu$_2$ will enhance our earlier picture \cite{3} obtained
through dc magnetization study on polycrystalline samples, the
additional results on the well known superconductor V$_3$Si will
probably indicate more universality of the concerned features.
We point out here that, although there exists report of PE in
V$_3$Si \cite{12}, our results (described below) will show the
existence of history effect and metastability associated with
the PE of V$_3$Si.

\section{Experimental}

The single crystal samples of CeRu$_2$ and V$_3$Si used in this
study are prepared by Dr. A. D. Huxley \cite{13} and Dr. A.
Menovsky \cite{14} respectively, and the details of the sample
preparation and characterization can be found in the references
13 and 14.  The residual resistivity ratio for the CeRu$_2$ and
V$_3$Si samples are 15 and 47 respectively \cite{13,14}.  The
electrical resistance measurements are performed using the
standard linear four probe method. A superconducting magnet and
cryostat system (Oxford Instruments, UK) is used to obtain the
required temperature (T) and magnetic field (H) environment. In
the configuration of our measurement the current (I$_M$) is
passed along the $<211>$ direction for the CeRu$_2$ sample and
the $<100>$ direction for the V$_3$Si sample.  In all the
measurements the direction of I$_M$ is kept perpendicular to H.
The T$_C$ for the CeRu$_2$ and V$_3$Si samples (obtained from
our zero field resistance measurements) are 6.1 and 16.5K
respectively.  The magnetic field dependence of the resistance
(R(H)) is measured,
\begin{enumerate}
\item on zero field cooling (ZFC) the sample 
to various T ($<$T$_C$) followed by an isothermal variation of H.
\item on field cooling (FC) the sample across T$_C$ 
in a fixed H to T ($<$T$_C$).  This is done for various H at each T.
\end{enumerate}

\section{Results and discussion}

For a sample of type-II superconductor with pinning, the
critical current (I$_C$) decreases with the increase in H and
goes to zero at the irreversibility field (H$_{irrv}
\leq$H$_{C2}$ ). 
However, for superconductors showing PE, I$_C$(H) shows a peak
or local maximum at an intermediate H value before finally going
to zero at H$_{irrv}$.  In an isothermal field variation, the
consequence of PE will show up in R(H) if the transport current
(I$_M$) used in the measurement is larger than I$_C$(H) for an
intermediate H regime but smaller than I$_C$(H) in the PE
regime.  In such a situation one would observe a flux flow
resistivity at fields H where I$_M >$I$_C$(H) but the resistance
will once again fall back to zero in the PE regime where I$_M<$
I$_C$(H).  Adjusting our measuring current I$_M$ appropriately,
we show in Fig. 1(a) and 1(b) the R vs H plots with distinct
signatures of PE for CeRu$_2$ and V$_3$Si.

(Although we have results obtained with various other measuring
currents at different temperatures, for the sake of clarity and
conciseness we shall use these two representative figures for
further discussions in the present work). The intermediate
flux-flow regime ( 0.8T$\leq$H$\leq$1.24T for CeRu$_2$ and
1.9T$\leq$H$\leq$3.47T for V$_3$Si) and the PE regime (
1.25T$\leq$H$\leq$1.37T for CeRu$_2$ and 3.48T$\leq$H$\leq$3.62T
for V$_3$Si) are very clear in these figures. As H approaches
H$_{irrv}$, the flux-flow resistivity starts appearing again
which ultimately leads to the normal state behaviour at
H$_{C2}$.
   
The PE regime for the CeRu$_2$ sample identified in our present
transport measurements agrees well with that obtained earlier
with the magnetic measurements \cite{2,3}. A comparison with the
sole magnetic measurement (to our knowledge) on V$_3$Si\cite{12}
leads to a similar conclusion.

We shall now focus on the field-history dependence of the
resistance in both CeRu$_2$ and V$_3$Si. The value of I$_M$ was
chosen for the two samples was such that the intermediate field
flux-flow was observed in both the systems when the field is
varied isothermally in the ZFC mode.  The measured resistance
is, however, found to be zero in the same field regime, when the
measurement is performed following the FC protocol (see Fig.2).
This clearly indicates that in this intermediate field regime
I$_C$(H) is greater than I$_M$ in the FC mode while I$_M $ is
greater than I$_C$(H) in the ZFC mode.  Thus, I$_C$(H) is higher
in the FC mode than in the ZFC mode.  A related history dependence of I$_C$ in polycrystalline
sample of CeRu$_2$ has earlier been reported by Dilley et al
\cite{9}. All these results of transport properties
measurements, we believe, are correlated to the anomalous FC
response observed in and around the PE regime of CeRu$_2$ in
various magnetic measurements \cite{2,3,15}.

However, (to our knowledge) no such report of history effects
exists for V$_3$Si either in magnetic properties or in transport
properties.  Similar field-history effects are well known in the
RFIM systems\cite{10,11,16}.  Most experimental information in
this regard has been obtained from various diluted
antiferromagnets in an applied magnetic field.  In ZFC mode, the
diluted antiferromagnet is cooled (in zero field) through the
zero-field Neel transition temperature. The resultant long range
magnetic order is preserved when an external magnetic field is
switched on at low temperatures.  This long range order,
however, gradually decreases on heating the sample to the high
temperature paramagnetic phase. However, on cooling back now
from the paramagnetic phase in the presence of the applied field
(i.e. in the FC mode), the sample develops a short range ordered
domain state \cite{10,11,16}.  The similarity with the vortex
state in CeRu$_2$ and V$_3$Si is apparent here, namely the
higher I$_C$ in the FC vortex state of these systems clearly
argues for a relatively more disordered FC vortex state. 

To draw the analogy further, we shall now deal with the
metastability of the FC state. It has been observed
experimentally that the FC state in the RFIMs are unstable to
field and temperature cycling below the phase boundary ; the FC
state tend to get back the long range order through such cycling
\cite{16,17}. To show the similar effects in the FC vortex state
of CeRu$_2$ and V$_3$Si we subject the sample to field cycling
after the initial field cooling experiment.  We have found that
the intermediate field zero resistance state is readily
destroyed by a subsequent field cycling through a small value
(of the order of few tens of mT) and the corresponding ZFC state
flux-flow resistance is recovered (see Fig. 2). Such
metastability is not observed in the low H regime ( H $<$ 0.8T
for CeRu$_2$ and H$<$1.9T for V$_3$Si) and inside the PE regime
( 1.25T$<$H$<$1.37T for CeRu$_2$ and 3.48T$<$H$<$3.62T for
V$_3$Si); the zero resistance state is quite stable to any field
cycling in these H-regime.

A continuous phase transition from an elastic vortex solid (or
Bragg-glass) to a plastic vortex solid (or Vortex-glass) has
been put forward as an explanation of the PE in various HTSC
materials \cite{18}.  However, any ZFC/FC history effect has not
been associated with the PE in these materials so far. Although
it is widely accepted that the PE in CeRu$_2$ indicates a
transition from a relatively ordered vortex solid to a
disordered vortex solid\cite{1,2,4,5,6,7,8,9} the exact nature
of this transition remains a matter of debate. We have earlier
suggested that the existence of a first order thermodynamic
phase transition and the associated supercooling can explain
various interesting features associated with PE in CeRu$_2$
including the history effects \cite{1,6,7}.  Here the history
effect is associated with the supercooling of the high field
high temperature disordered vortex state (with enhanced pinning)
across the transition line.  This picture gains further weight
by our recent theoretical argument that the range of
supercooling will be more while varying the temperature than
while varying the field \cite{19}. The experimental support in
this regard already existed in magnetic studies of CeRu$_2$, and
further support is obtained through our present transport
measurements. There is a finite hysteresis in the field
variation of the resistance in CeRu$_2$ between the ascending
and the descending field cycle (see inset of Fig. 1(a)). Such a
hysteresis in the isothermal field variation of R(H), however,
is not very distinct for V$_3$Si.  As shown in Fig.2, the path
dependence of R(H) is very clear for both CeRu$_2$ and V$_3$Si
when the vortex state prepared through the FC protocol. In spite
of all these arguments, we must point out here that a definite
microscopic experimental evidence in support of (or against) a
first order phase transition in CeRu$_2$ is yet to appear.  The
analogy with the RFIM systems continues here, since the question
regarding the underlying phase transition is yet to be settled
in the RFIM systems also \cite{11,20}.

\section{Conclusion}
We have presented results of transport measurements in CeRu$_2$
which, in conjunction with our earlier magnetic measurements
\cite{3}, show clear analogy between the vortex state of CeRu$_2$
around the PE regime and the RFIM systems. Our study shows that
like in RFIM systems, the FC vortex state in CeRu$_2$ is
relatively more disordered and metastable in character. In
addition, we have shown here the existence of the same features
in the well known superconductor V$_3$Si.

\newpage
\begin{center}\large \bf Figure Captions \end{center}

\begin{itemize}

\item[Fig. 1] (a) Resistance (R) vs field (H) plot for CeRu$_2$ 
obtained in the ZFC mode at 5K with I$_M$ = 100 mA. Open squares
(open triangles) denote the data points in the ascending
(descending) H cycle.  The inset shows the hysteresis in R(H)
between the ascending and the descending H cycle, at the onset
of the PE regime. (b) Resistance (R) vs Field (H) plot for
V$_3$Si at 14.5K with I$_M$ = 85 mA.

\item[Fig. 2] (a) Metastable behaviour of R(H) of CeRu$_2$
obtained in the FC mode at 5K with I$_M$ = 100 mA. Solid circles
denote R(H) values obtained after field cooling in various H
values. Filled triangles denote R(H) values obtained after a
field cycling of maximum 25 mT subsequent to the first FC
measurement at the corresponding H. Solid and dashed lines
denote the R(H) obtained in the isothermal ZFC mode in the
ascending and descending H respectively (same as in Fig. 1(a)).
(b) Metastable behaviour of R(H) of V$_3$Si obtained in the FC
mode at 14.5K with I$_M$ = 85 mA. Solid circles denote R(H)
values obtained after field cooling in various H values. Filled
triangles denote R(H) values obtained after a field cycling of
maximum 50 mT subsequent to the first FC measurement at the
corresponding H. Solid and dashed lines denote the R(H) obtained
in the isothermal ZFC mode in the ascending and descending H
respectively (same as in Fig. 1(b)).

\end{itemize}

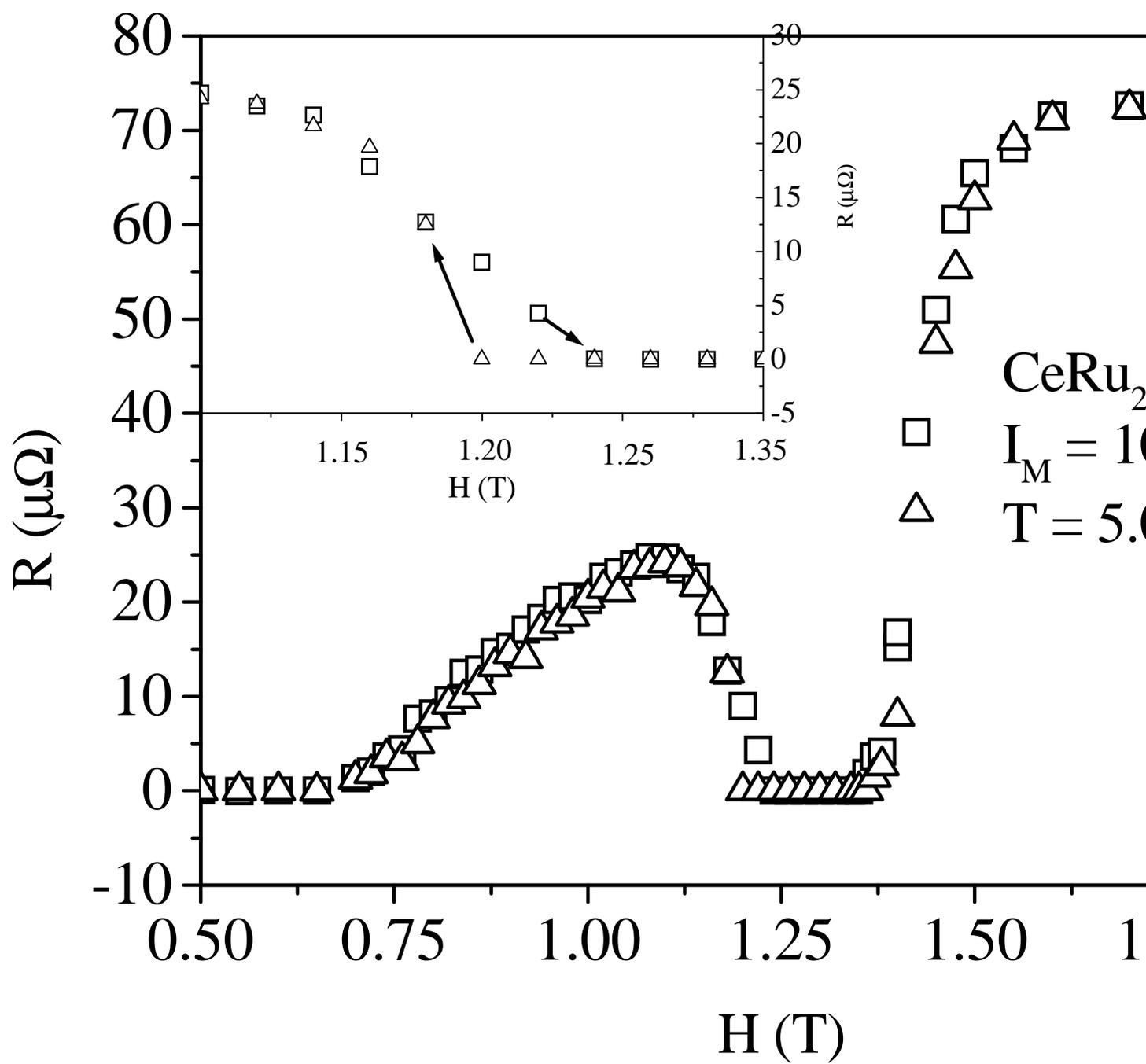
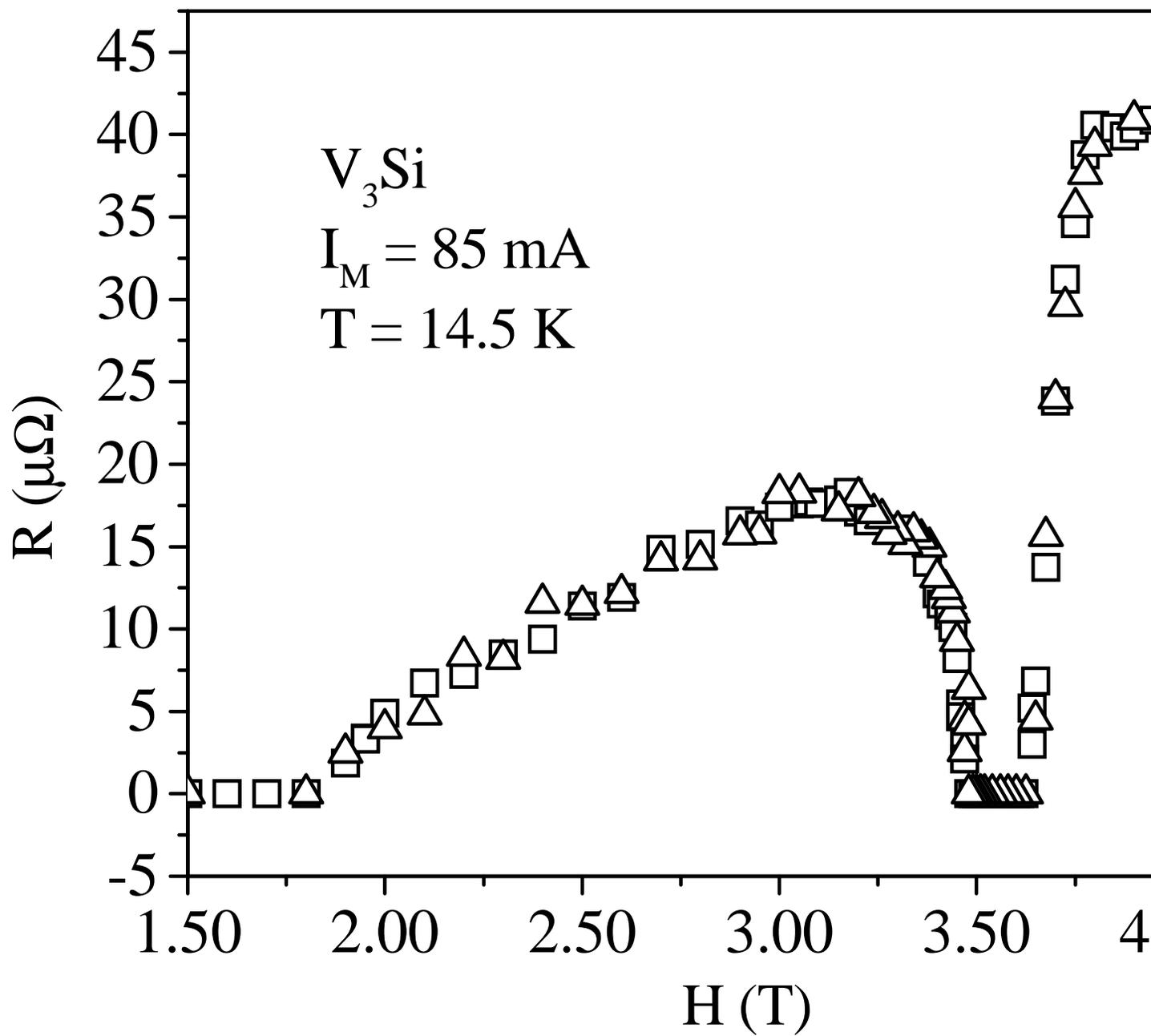
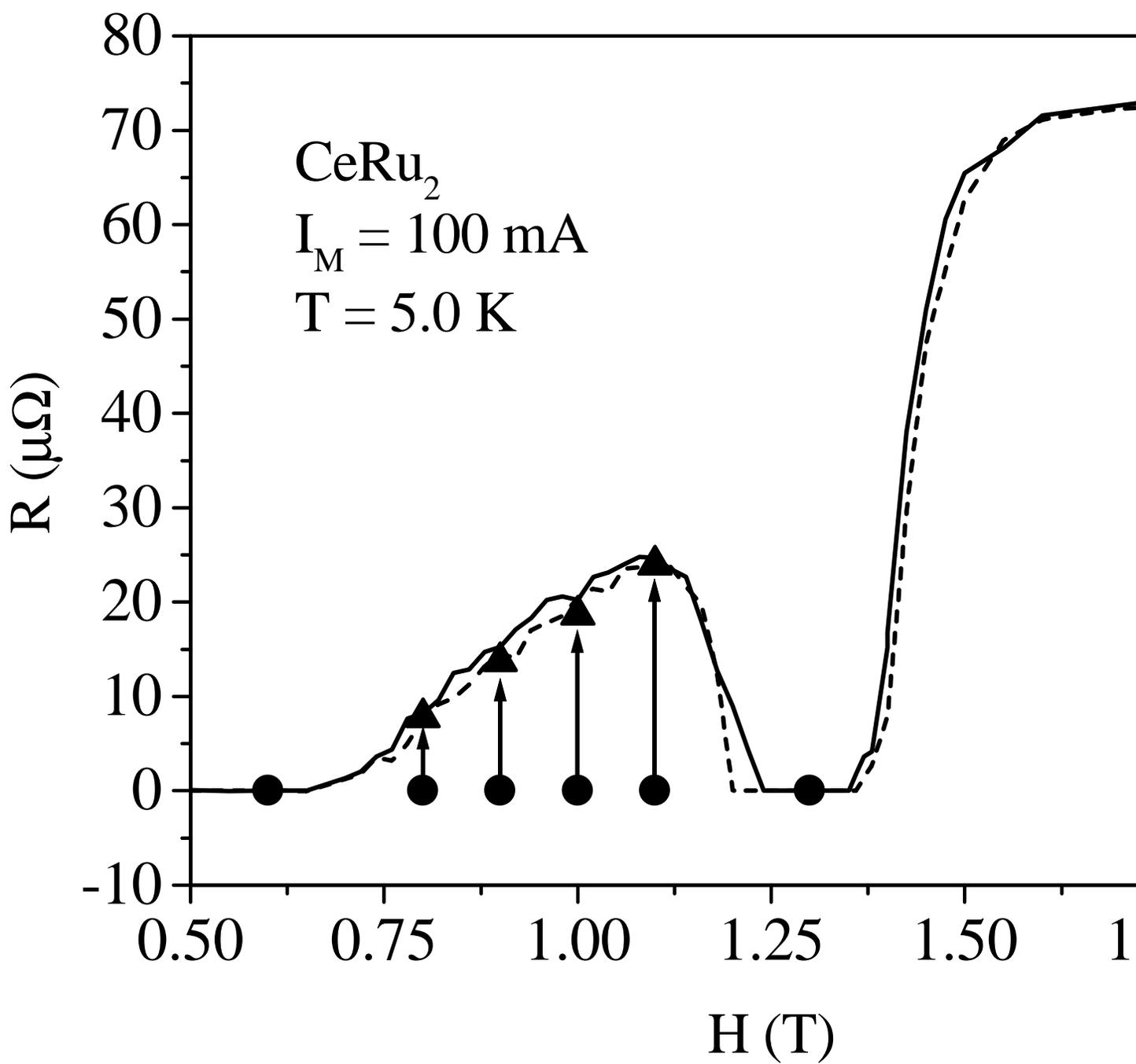
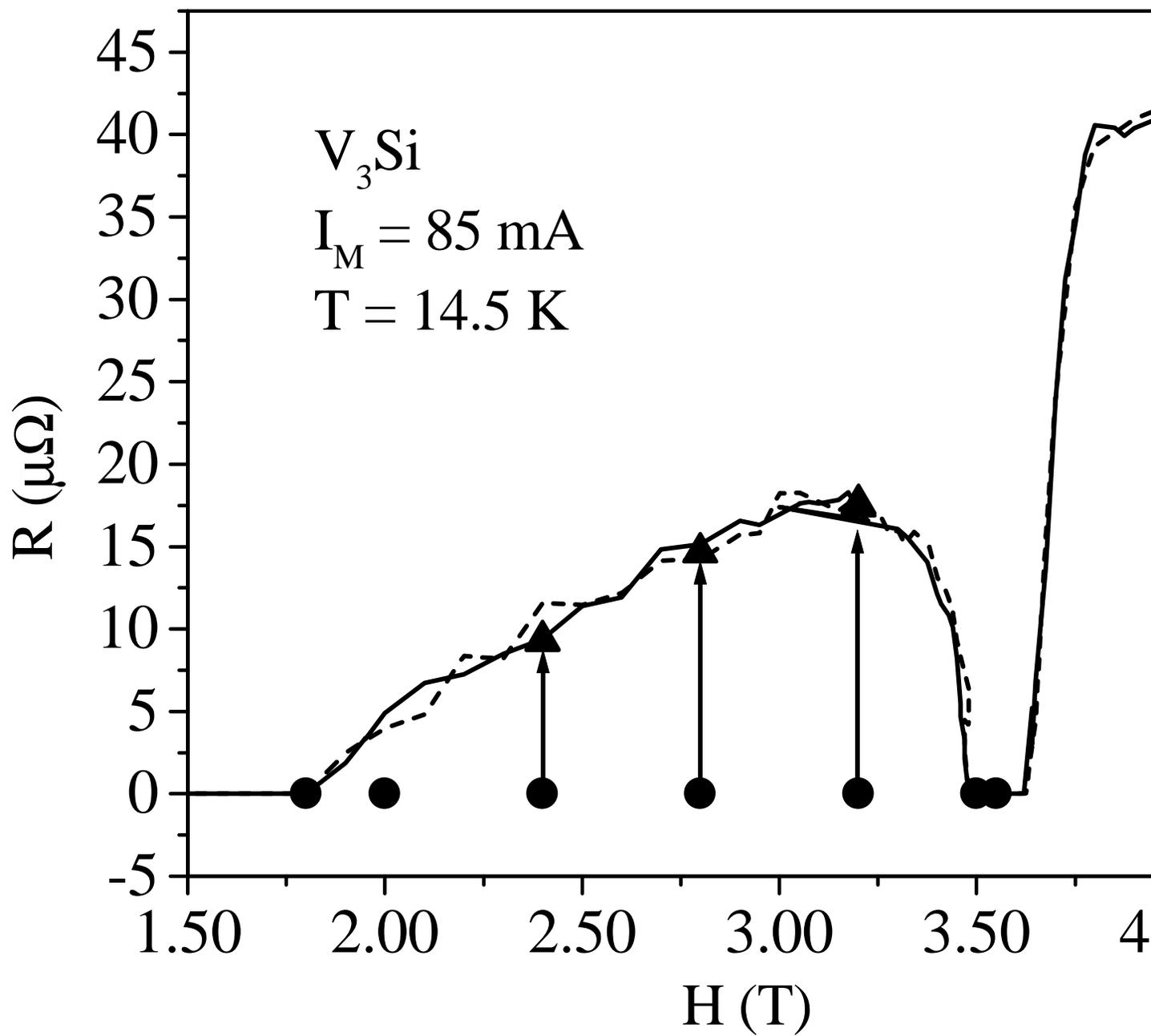
\end{document}